\documentclass[twocolumn,superscriptaddress, a4paper,longbibliography]{revtex4-1}
\usepackage[T1]{fontenc}
\usepackage[english]{babel}
\usepackage{siunitx}
\usepackage[utf8]{inputenc}
\usepackage{url}
\usepackage{dsfont}
\usepackage{bbm}
\usepackage{xcolor}
\usepackage{nicefrac}
\usepackage{setspace}
\usepackage{float}
\usepackage{comment}
\usepackage{graphicx}
\usepackage{ulem}
\usepackage[colorlinks=true, citecolor=blue,urlcolor=blue,linkcolor=blue,filecolor=black]{hyperref}
\usepackage{amsthm}
\usepackage{mathtools}
\usepackage{physics}
\usepackage{graphicx}
\usepackage[left=23mm,right=13mm,top=35mm,columnsep=15pt]{geometry} 
\usepackage{adjustbox}
\usepackage{placeins}

\usepackage{lipsum}
\usepackage{csquotes}

\newcommand{\beq}{\begin{equation}}
\newcommand{\eeq}{\end{equation}}
\renewcommand{\emph}{\textit}

\usepackage{color,soul}

\def\ket #1{\vert #1\rangle}

\usepackage{amsmath,amssymb}

\begin{document}

\title{Stable, low-error and calibration-free polarization encoder for free-space quantum~communication}

\author{Marco Avesani}
\thanks{These authors contributed equally to this work.}
\affiliation{Dipartimento di Ingegneria dell'Informazione, Universit\`a degli Studi di Padova, Via Gradenigo 6B - 35131 Padova, Italia}

\author{Costantino Agnesi}
\thanks{These authors contributed equally to this work.}
\affiliation{Dipartimento di Ingegneria dell'Informazione, Universit\`a degli Studi di Padova, Via Gradenigo 6B - 35131 Padova, Italia}
\thanks{These authors contributed equally to this work.}

\author{Andrea Stanco}
\affiliation{Dipartimento di Ingegneria dell'Informazione, Universit\`a degli Studi di Padova, Via Gradenigo 6B - 35131 Padova, Italia}

\author{Giuseppe Vallone}
\affiliation{Dipartimento di Ingegneria dell'Informazione, Universit\`a degli Studi di Padova, Via Gradenigo 6B - 35131 Padova, Italia}
\affiliation{Dipartimento di Fisica e Astronomia, Università degli Studi di Padova, via Marzolo 8, 35131 Padova, Italy}
\affiliation{Istituto di Fotonica e Nanotecnologie - CNR, Via Trasea 7 - 35131 Padova, Italia}

\author{Paolo Villoresi}
\affiliation{Dipartimento di Ingegneria dell'Informazione, Universit\`a degli Studi di Padova, Via Gradenigo 6B - 35131 Padova, Italia}
\affiliation{Istituto di Fotonica e Nanotecnologie - CNR, Via Trasea 7 - 35131 Padova, Italia}

\thanks{These authors contributed equally to this work.}

\begin{abstract}
Polarization-encoded free-space Quantum Communication requires a quantum state source featuring fast polarization modulation, long-term stability and a low intrinsic error rate.
Here we present a source based on a Sagnac interferometer and composed of polarization maintaining fibers, a fiber polarization beam splitter and an electro-optic phase modulator. The system generates predetermined polarization states with a fixed reference frame in free-space that does not require calibration neither at the transmitter nor at the receiver. In this way we achieve long-term stability and low error rates.
A proof-of-concept experiment is also reported, demonstrating a Quantum Bit Error Rate lower than 0.2\% for several hours without any active recalibration of the devices. 
\end{abstract}

\maketitle

\section{Introduction}
Quantum key distribution (QKD) allows two spatially distant parties to generate, through the exchange of qubits encoded into single photons, a secure and private cryptographic key. 
The security and privacy of the keys generated by QKD is guaranteed by the principles of quantum mechanics.

Practical implementations of QKD require rapid modulation of the degrees of freedom of the photons used to encode the qubits. 
One of the most exploited encoding, in particular for free-space application, is represented by polarization.
A quantum source featuring long-term stability and a low intrinsic Quantum Bit Error rate (QBER) is crucial for 
 free-space ~\cite{avesani2019daylight,Liao2017daylight}, and in particular satellite~\cite{Liao2017_Sat,Vallone2015prl}
 QKD application.
Nonetheless, most of the methods used so far 
to modulate the photon's polarization were characterized by a low temporal stability and a limited contrast in the modulation due to source imperfections. Moreover, some implementation could lead to side-channel attacks, undermining the security of the QKD protocol.

The simplest solution is the use of different lasers, one for each state required by the QKD protocol. This approach has been used in several works 
\cite{Liao2017_Day,bacco13, Vest2015}
and in particular in the recent satellite-to-ground demonstration~\cite{Liao2017_Sat}.
This solution could be demanding from the economic and energetic point of view as it requires four independent lasers,  temperature controllers and current generators. 
More importantly, if the lasers are not properly calibrated and controlled, the generated states can be distinguished and the implementation could be vulnerable to attacks such as those described recently in~\cite{Lee2019}.

In order to prevent the above attacks a single laser source with an active polarization modulator
is typically used.
A common technique is based on the use of 
birefringent phase modulators in an in-line configuration~\cite{jofre2010,Grunenfelder2018}.
In this scheme, the photons are injected in the phase modulator (typically based on LiNBO$_3$ crystals) with a polarization state that is diagonal with respect to the optical axis of the modulator's crystal.
By applying a $V_{\rm RF}$ voltage, the ordinary and extraordinary refractive indices of the crystal vary independently, allowing to control the relative phase between the $\ket H$ and $\ket V$ polarization.

\begin{figure*}[t]
\centering
\includegraphics[width=0.92\linewidth]{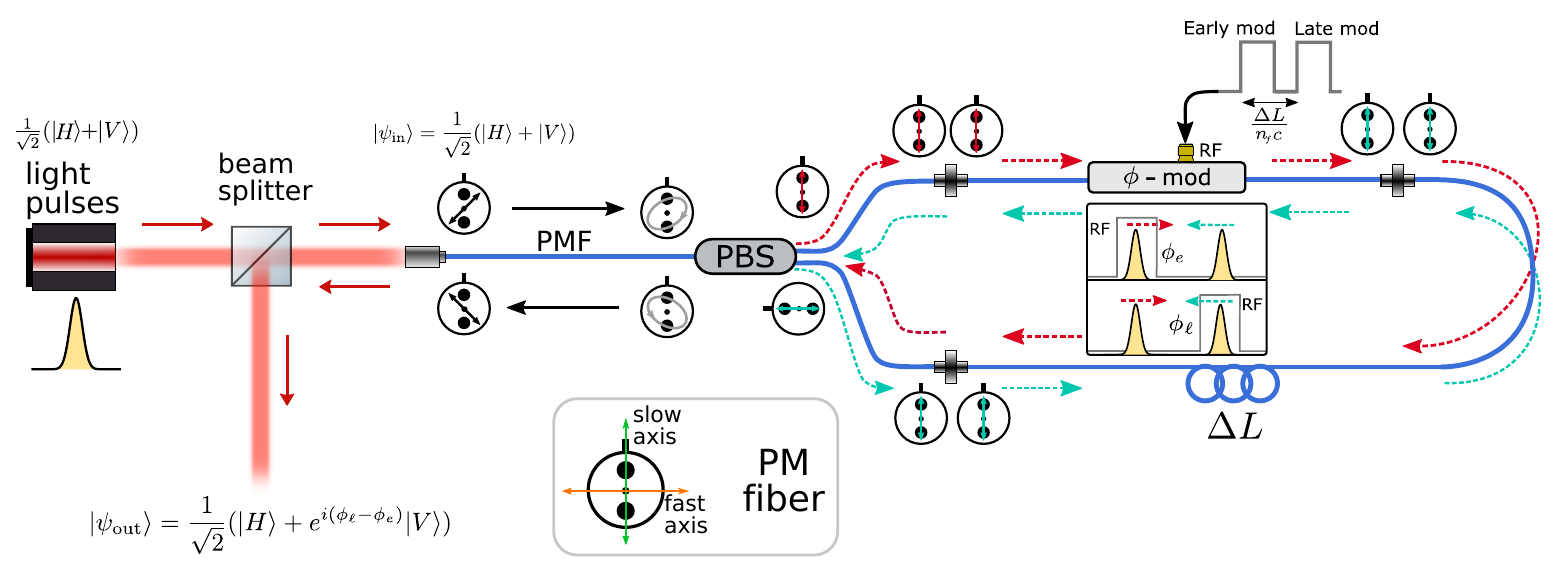}
\caption{Scheme of the proposed polarization modulator. The free-space beam-splitter is used to separate incoming from outcoming light. A Sagnac interferometer including a fiber polarizing beam splitter, a phase modulator and polarization Maintaining fibers (indicated in blue) is used to modulate the polarization state. We always consider a left-handed reference frame whose $z$ axis is directed toward the photon's propagation direction.}
\label{fig:setup}
\end{figure*}

This solution presents several disadvantages. First of all, the polarization modulation is highly sensible to temperature and bias voltage drift and requires active stabilization. A second issue is related to the polarization mode dispersion (PMD) induced by the birefringence, which reduces the degree of polarization for short pulses, decreasing the modulation performance. Thirdly, high voltages are required to modules to induce the required polarization modulation.
Lastly, modulators able to support both polarization are not always commercially available, especially outside the C band.

To avoid temperature and bias voltage stabilization, it is possible to use the birefringent modulator in a two-pass scheme using a Faraday mirror~\cite{Martinez2009}.
However, this scheme still requires non-standard components and high voltages to induce the necessary modulation. Moreover, PMD is also not completely compensated, allowing only modest modulation performances~\cite{Martinez2009}.

To solve the problems of the polarization modulators mentioned above, we have recently proposed the POGNAC~\cite{Agnesi2019} and exploited it in a complete QKD experiment featuring the lowest intrinsic QBER  reported to date~\cite{Agnesi2019b}.
The POGNAC uses an optical circulator and a single mode fiber (SMF) to guide the photons into a fiber Polarization Beam Splitter (PBS) that represents the beginning of a Sagnac interferometer that includes a standard phase modulator ($\phi$-mod). 
A polarization controller (PC) is used to obtain at the input of the fiber PBS a balanced overlap of the horizontal $\ket H$ and vertical $\ket V$ polarization modes. The photons exit back from the fiber PBS with a polarization state given by $\ket{\psi}=\frac{1}{\sqrt2}(\ket{H}+e^{i\phi}\ket V)$, with $\phi$ determined by the $\phi$-mod, and then are directed toward the exit port of the circulator. 
Some similar schemes require the introduction of a Faraday rotator in the Sagnac loop~\cite{Jiang2019,Li2019}.
 The Sagnac loop geometry has also been exploited for polarization-independent phase modulation~\cite{Wang2018} as well as intensity modulation~\cite{Roberts2018}.

The Sagnac-based polarization modulators described above have some limitations related to the presence of the single-mode fiber for optical input and output. Indeed, the fiber requires that the device is manually calibrated via the PC to obtain a balanced overlap of the horizontal and vertical polarization modes at the input of the PBS. This is a disadvantage in the development of autonomous systems. Moreover, temperature drifts require (slow) recalibration of the device.
Moreover, at the exit of the Sagnac interferometer the single mode fiber that guides the light from the PBS to the circulator introduces a unitary transformation of the polarization state that is not known a priori. This complicates the implementation of QKD protocols in free-space (and in particular in satellite-to-ground QKD) as it introduces a further calibration phase at the receiver. 
Furthermore, temperature variations and mechanical stress of the fiber slowly change the corresponding unitary transformation. 

The issue related to the polarization calibration at the input has been solved in~\cite{Li2019a} where a polarization maintaining fiber (PMF)  with optical axis oriented at $45^\circ$ is used at the input of the Sagnac loop. However, the use of SMF at the output still requires a calibration phase at the receiver since the emitted state are generated up to an unknown unitary transformation.

In this manuscript we present iPOGNAC, a polarization encoder that solves all the previously mentioned issues with a simple scheme.
The iPOGNAC, generates predetermined polarization states with a fixed reference frame in free-space that does not require calibration neither at the transmitter nor at the receiver, featuring a long-term stability and a low intrinsic QBER.
We think that this polarization encoder provides a simple and effective solution for free-space quantum communication.

\section{Proposed scheme }
The scheme of the iPOGNAC can be seen in Fig.~\ref{fig:setup}.
Collimated laser pulses propagating in free-space with diagonal polarization $\ket{D}= \left( \ket{H} + \ket{V} \right)/\sqrt{2}$ (considering a left-handed reference frame whose $z$ axis is directed toward the photon's propagation direction) enter the setup. This polarization state can be generated in several ways, for example by using a polarizer, by using a Half Wave Plate (HWP), or by physically rotating the collimator if the light is delivered by a laser emitting light coupled to the slow (or fast) axis of a PMF.  
It is important to note that any state that is a balanced superposition of $\ket{H}$ and $\ket{V}$, i.e., laying in the equator of the Bloch sphere, is equally suitable.

The light pulses then impinge a free-space Beam Splitter (BS). The reflected light is discarded, while the transmitted pulses encounter a collimating lens that couples the light into a PMF with slow axis aligned to the vertical direction. Since PMFs are designed to be birefringent, the polarization modes $\ket{H}$ and $\ket{V}$ experience different indexes of refraction and the polarization state at the end of the PMF is changed to an elliptical polarization $
\ket{\psi_1}= \left( \ket{H} +  e^{i \delta} \ket{V} \right)/\sqrt{2}$.
The phase $\delta$ depends on the length of the fibre, the difference between the fast and slow indexes of refraction of the PMF, temperature and mechanical strain applied to the PMF. Interestingly enough, for laser pulses with short coherence times the phase $\delta$ can be large enough to effectively depolarize the light.

After propagating in the PMF, the light encounters a fiber-based Polarization Beam Splitter (PBS).
From this point the setup is identical to the original POGNAC~\cite{Agnesi2019}, which allows us to inherit its high intrinsic stability and the capability of modulating polarization exploiting a $\phi$-mod.
The light is split into orthogonal linear polarizations by the fiber PBS.
It is important to note that each of the polarized beams exiting from the PBS is aligned to the slow axis of a PM fiber. This effectively maps the polarization degree of freedom onto the optical path of the photons, with the polarized light traveling along only the slow axis of the PM fibers of both PBS exit ports.
This is the standard behavior of fiber-based polarization beam combiners and splitters.

This PBS marks the beginning of the Sagnac interferometer, fully implemented with PM fibers.
The vertically polarized component travels in the clockwise direction (CW) while the horizontally polarized component travels in the counter-clockwise direction (CCW).
In the CW direction a $\phi$-mod is first encountered introducing an "Early" phase $\phi_e$ to the CW propagating light pulse.
A PMF delay line is then encountered, after which the CW light pulse impinges once again on the PBS.
The CW propagating light exits the Sagnac interferometer with horizontal polarization.
In the reverse direction, the CCW first encounters the PMF delay line such that it arrives on the modulator later than the CW pulse. 
Then, the $\phi$-mod introduces a "Late" phase $\phi_\ell$ to the CCW propagating light pulse. 
Lastly, the CCW light pulse impinges once again on the PBS, exiting the Sagnac interferometer with vertical polarization.

Since inside the PMF Sagnac interferometer, both the CW and CCW travel along the fast axis of the PM fiber, no polarization mode dispersion is observed and a single polarization mode propagates in the $\phi$-mod. 

The polarization state emerging from the PBS is thus given by 
\begin{equation}
 \ket{\psi^{\phi_e,\phi_\ell}_2}  = \frac{1}{\sqrt{2}} \left( \ket{H} - 
 e^{i(\phi_\ell- \phi_e- \delta)} \ket{V} \right)\,.
\end{equation}
It is important to note that the light that entered the Sagnac loop with $\ket{H}$ polarization emerge with $\ket{V}$ polarization, and vice versa. This has the effect of changing the sign of the phase $\delta$ and introducing an extra $\pi$ shift. Indeed, by using a left-handed reference frame whose $z$ axis that is always directed toward the photon's propagation direction, the PBS and the Sagnac loop without modulation implement an $i\sigma_y$ operation, namely
$\ket H\rightarrow-\ket V$ and 
$\ket V\rightarrow\ket H$.

After exiting the fiber PBS that closes the Sagnac loop, the light propagates backwards through the PMF, which adds an additional phase $\delta$ between the propagating polarization modes. Interestingly enough this additional $\delta$ phase compensates the phase accumulated by the forward propagation in the PMF. 
It is worth noting that such phase $\delta$ is compensated also for laser pulses with short coherence times, even when the delay between the two modes is larger than the coherence length.
This can be easily understood since the $\ket{H}$ ($\ket{V}$) polarized light that first traveled along the fast  (slow) axis, emerges as $\ket{V}$ ($\ket{H}$) after the Sagnac loop and travels back along the slow (fast) axis.  

\begin{figure*}[t!]
\centering  \includegraphics[width=\linewidth]{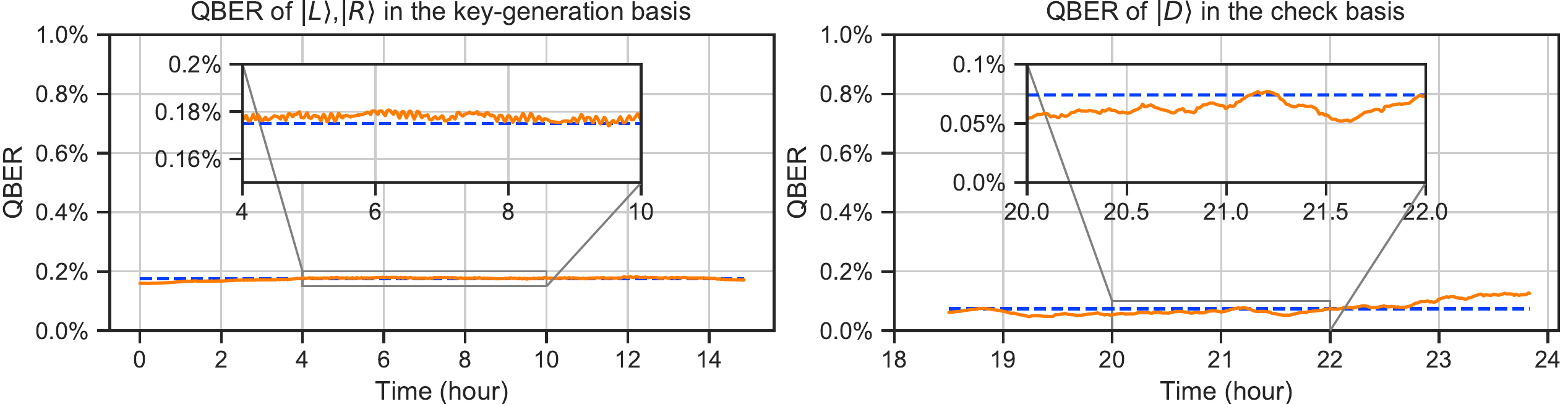}
\caption{Experimental QBER as a function of time measured from $t_0$=18:30 of 12 Sept. 2019. The left panel shows the QBER in key-generation basis, were an average value of $\mathcal{Q_K} = 0.175\%\pm0.005\%$ was obtained. The right panel shows the QBER in check basis, were an average value of $\mathcal{Q_C} = 0.07\%\pm0.02\%$ was obtained. The time between 14.5 and 18.5 hour was needed for the helium cycle of the SNSPD detectors. During this time, no operation on the source was performed.
}
\label{fig:QBER}
\end{figure*}

The collimating lens then emits the light back into free space. 
The  light then impinges once more the free-space BS and the transmitted light is discarded. 
On the other hand, the exit on the reflection port of the free-space BS represents the output of the iPOGNAC polarization modulator. Due to an extra $\pi$ shift due to the reflection,  the polarization state of light at the output is described by
\begin{equation}
 \ket{\psi^{\phi_e,\phi_\ell}_\mathrm{out}}  = \frac{1}{\sqrt{2}} \left( \ket{H} + 
 e^{i(\phi_\ell - \phi_e)} \ket{V} \right)\,.
\end{equation}

Since inside the Sagnac loop the CW pulse anticipates the arrival of the CCW pulse on the $\phi$-mod by a factor  $\frac{\Delta L}{n_f c}$ (where $n_f$ is the index of refraction of the slow axis of the PMF and $c$ the velocity of light), by carefully timing the applied voltage on the $\phi$-mod, the polarization state $\ket{\psi_\mathrm{out}}$ can be modulated.

Indeed, by changing the phases $\phi_e$ and $\phi_\ell$ any state lying on the equator of the Bloch sphere can be generated. For QKD application, a single voltage level $V_{\pi/2}$ (corresponding to the voltage required to introduce a phase shift of $\pi/2$ in the $\phi$-mod) and different timing 
are sufficient to generate the states $\ket{D}$, $\ket{L}=\frac{1}{\sqrt{2}} \left( \ket{H} + i \ket{V} \right)$ and $\ket{R}=\frac{1}{\sqrt{2}} \left( \ket{H} - i \ket{V} \right)$, allowing the implementation of the 3-state version of the BB84 protocol~\cite{Grunenfelder2018}. 
In particular, if no voltage (or equal voltage) is applied to the CW and CCW pulses, the polarization state remains unchanged, i.e.
$\ket{\psi_\mathrm{out}^{0,0}} = \ket{D}$.
If $V_{\pi/2}$ voltage is applied to the CW pulse and no voltage is applied to the CCW pulse, the output state becomes  
$\ket{\psi_\mathrm{out}^{\frac{\pi}{2},0}} = \ket{L}$.  
Alternatively, if no voltage is applied to the CW pulse and $V_{\pi/2}$ voltage is applied to the CCW pulse, the output state becomes  
$\ket{\psi_\mathrm{out}^{0,\frac{\pi}{2}}} = \ket{R}$.
We note that, by using a generic input state of the form $\alpha\ket H+\beta \ket V$,
the output states 
$\beta\ket{H} + 
 \alpha e^{i(\phi_\ell - \phi_e)} \ket{V})$
 can be generated.

It is important to stress that the proposed scheme requires only standard components, reducing construction costs and allowing the scheme to be used at different wavelengths, such as 800nm, 1064nm or 1550nm.

\section{Proof-of-concept experiment}

We performed a proof-of-concept experiment of the iPOGNAC. As a light source we used a gain-switched distributed feedback (DFB) laser which emitted a 50~MHz stream of phase-randomized pulses with 270~ps full-width-at-half-maximum duration at 1550 nm. To generate the the $\ket{D}$ polarization state a Glan-Thompson polarizer was used. The iPOGNAC was assembled using standard Commercial Off-The-Shelf (COTS) components exclusively. 
The electronic signals used to drive the iPOGNAC were generated by a Field Programmable Gate Array (FPGA) (Zynq-7020) mounted on a dedicated board (ZedBoard by Avnet). The FPGA was capable of triggering the DFB laser and to produce the "Early" and "Late" modulation pulses with specific time delays in order to address the $\phi_e$ and $\phi_\ell$ phases (all the signals were properly amplified by RF amplifiers). The output of the iPOGNAC was then attenuated to the single-photon level using attenuators and directed to a free-space polarization analyzed comprised of a Quarter Wave Plate (QWP), a HWP and a PBS.
The output ports of the polarization analyzer were then coupled to single-mode fibers which directed the light to two superconductive nanowire single-photon detectors (SNSPD) with detection efficiencies around 85\%.

In order to test the long-term stability of our system, a fixed sequence composed by the three polarization states $\{\ket{L},\ket{R},\ket{D}\}$ is prepared by the iPOGNAC.
We note that those three states are sufficient to realize the three-state BB84 protocol presented in~\cite{Rusca2018}.
The states are then measured by a free-space measurement station projecting the states in the key-generation basis $\mathcal K=\{\ket{L},\ket{R}\}$
or in the check basis
$\mathcal C=\{\ket{D},\ket{A}\}$ with
$\ket A=1/\sqrt{2}(\ket H-\ket V)$. The selection of the basis is performed by acting on the QWP placed before the PBS of the measurement station. The timestamps of the photo-detection events are digitized by time-to-digital converted and then streamed to the PC for the post-processing analysis.
We performed two acquisitions: the first one, lasting almost 15 hours and acquired during the night, measures the QBER in the key-generation basis $\{\ket{L},\ket{R} \}$ and is presented in the left panel of Fig \ref{fig:QBER}. In this configuration we measured an average QBER for the $\ket{L},\ket{R}$ states $\mathcal{Q_K}$ of $0.175\% \pm 0.07\%$. 

After that, the helium cycle of the SNSPD was performed in order to keep the detectors at the operating temperature. The cycle lasted 4 hours and during this time no operation on the source was performed.
Then, by acting on the QWP at the measurement station, we aligned the measurement station along the check-basis $\mathcal{C}$ and we performed another acquisition lasting 5 hours and half, during the day. The data are presented in the right panel of Fig \ref{fig:QBER}. We measured an average QBER for the $\ket{D}$ state of $\mathcal{Q_C}=0.07\%\pm0.02\%$.

In both cases, the data shown in Fig. \ref{fig:QBER} confirm the excellent performances of the iPOGNAC in terms of QBER and long-term stability, making it optimal solution as polarization encoder for classical and quantum free-space communications.

The small and slow residual drifts visible in Fig \ref{fig:QBER} can be attributed to the temperature variations of the measurement apparatus. In fact, the lab was not temperature stabilized and experiences variation of several degrees during the day. These variations can induce small polarization rotations to the photons travelling in the single mode fibers connected to the SNSPD. Since the efficiency of the SNSPD is polarization-dependent, these rotations can induce small drifts in the measured QBER.

\section{Conclusions}

In conclusion, we proposed and tested a novel polarization encoder, called iPOGNAC, characterized by a low intrinsic error and high temporal stability. Moreover, the encoder can generate predetermined polarization states with a fixed reference frame in free-space.

We experimentally implemented the proposed protocol, using only COTS components, obtaining a low intrinsic QBER $<0.2\%$ and long-term stability measured over 24 hours. Indeed, these performances are well-suited for practical QKD systems.

In particular, the iPOGNAC is a promising solution for quantum communication with satellites or moving objects for at least two reasons. Firstly, its ability of generating polarization states that are fixed with respect to  the transmitter's reference frame eliminates the need of calibration between the transmitter and the receiver. Secondly, given its free-space output, it can be easily interfaced with a telescope. 
Moreover, the proposed source, further miniaturized with micro-optic technology, can be also exploited in fiber-based QKD with the requirement of a receiver calibration. 
Finally, the iPOGNAC could be also used at the receiver side in a QKD application, to rapidly switch between measurement basis, or for classical communication exploiting polarization modulation~\cite{Benedetto1992}.

The device here presented is object of the Italian Patent Application No. 102019000019373 filed on 21.10.2019
\cite{vallone2019}.

\vspace{1cm}
{\it Note added} - After the completion of the present manuscript we became aware of a similar proposal presented in~\cite{Xu2020}.

\begin{acknowledgements}
Part of this work was supported by MIUR (Italian Minister for Education)
under the initiative ``Departments of Excellence'' (Law 232/2016)
\end{acknowledgements}

\end{document}